\documentclass[aps,onecolumn,groupedaddress,nofootinbib,notitlepage]{revtex4-1}

\usepackage{mathtools}
\usepackage{amsfonts}
\usepackage[format=plain,justification=centerlast,width=0.8\textwidth]{caption}
\usepackage{enumerate}
\usepackage[utf8]{inputenc}
\usepackage{rotating}
\usepackage{enumitem}
\usepackage{algorithm}
\usepackage{algorithmic}
\usepackage[english]{babel}
\usepackage{calc}

\newcommand{\abs}[1]{ \left| #1 \right| }
\newcommand{\bra}[1]{\langle #1|}
\newcommand{\ket}[1]{|#1\rangle}

\begin{document}

\title{Exact simulation of coined quantum walks with the continuous-time model}

\author{Pascal Philipp}
\affiliation{National Laboratory of Scientific Computing, Petr\'opolis, RJ, Brazil}
\author{Renato Portugal}
\affiliation{National Laboratory of Scientific Computing, Petr\'opolis, RJ, Brazil}

\date{\today}

\begin{abstract}
The connection between coined and continuous-time quantum walk models has been addressed in a number of papers. In most of those studies, the continuous-time model is derived from coined quantum walks by employing dimensional reduction and taking appropriate limits. In this work, we produce the evolution of a coined quantum walk on a generic graph using a continuous-time quantum walk on a larger graph. In addition to expanding the underlying structure, we also have to switch on and off edges during the continuous-time evolution to accommodate the alternation between the shift and coin operators from the coined model. In one particular case, the connection is very natural, and the continuous-time quantum walk that simulates the coined quantum walk is driven by the graph Laplacian on the dynamically changing expanded graph.
\end{abstract}


\maketitle

\section{Introduction}

Quantum walks (QWs) \cite{PhysRevA.48.1687,Aharonov:2001:QWG:380752.380758,PhysRevA.58.915,doi:10.1080/00107151031000110776} are the quantum analogue of classical random walks (RWs) \cite{hughes1995random,lawler2010random}, and one of the main building blocks for quantum algorithms \cite{doi:10.1142/S0219749903000383}. An important example, which demonstrates the impressive potential of quantum computing, is given by quantum search algorithms \cite{grover1996,PhysRevA.70.022314,Ambainis:2005:CMQ:1070432.1070590,PhysRevA.92.032320,PhysRevLett.116.100501}. While the lower bound on the time complexity of classical search algorithms on unsorted databases is linear in size, quantum search algorithms often have, depending on the structure of the database, time complexity sublinear in size. Another example for the improvement of classical algorithms is the element distinctness problem \cite{doi:10.1142/S0219749903000383,Childs2009}. The development of quantum PageRank algorithms \cite{paparo2013} and the simulation of neutrino oscillations \cite{2016arXiv160404233M} provide further examples for the wide range of applications of QWs.

For both RWs and QWs, there are discrete-time and continuous-time formulations. For RWs, the relation between these two formulations is very close, both formally and with regards to the behavior over time. In fact, the transition matrix of a continuous-time RW is obtained from the discrete-time transition matrix either via limits or by using a Poisson process to transfer the time variable into a continuous domain\footnote{Let $n_t$ be a Poisson random variable and denote the expected value by $\langle\:\cdot\:\rangle$. For the continuous-time transition matrix $M_{CT}(t)$ and the discrete-time transition matrix $M_{DT}(k)$ of the RW, we have
\[                                                                                                                                                                                                                                                                                                                                                                                                                                                                                                                                 M_{CT}(t) = \langle M_{DT}(n_t) \rangle = \sum_{k=0}^\infty P(n_t=k)M_{DT}(k) = \text{exp}[t(M_{DT}-I)]
.\]
}.

The relation between discrete-time QWs (DTQWs) and continuous-time QWs (CTQWs) is less straightforward. The dimension of the Hilbert space in which a CTQW~\cite{PhysRevA.58.915} takes place is equal to the number of vertices of the underlying graph. A first stepping stone for linking DTQWs and CTQWs comes from the need for additional degrees of freedom in the discrete-time model, which is indirectly stated by the no-go lemma \cite{Meyer1996337}. For instance, the DTQW models in Refs.~\cite{PhysRevA.48.1687,Aharonov:2001:QWG:380752.380758,doi:10.1080/00107151031000110776} introduce those extra degrees of freedom via ``quantum coins''. More precisely, the no-go lemma states that on $d$-dimensional lattices, there exist no nontrivial, homogeneous, scalar unitary cellular automata. For example, one can generate nontrivial but inhomogeneous evolution by alternating the action of local operators. The staggered QW model~\cite{Portugal2016staggered,2016arXiv160302210P} is an example of this class, and it has no internal space. If the evolution is driven by the application of two local operators, say $U_1$ and $U_2$, then setting $U=U_2U_1$ would restore homogeneity but break locality by allowing the walker to take two steps per time unit. However, in the context of QWs \emph{on graphs}, it seems preferable to respect locality. In coined models, there are two evolution operators involved as well, but they can be combined without violating locality since the coin space is not considered to be spatial.

It is known that DTQWs and CTQWs have the same or similar features, such as the same spreading rate on lattices, similar probability distributions, and almost the same asymptotic behavior for large time. Connections between DTQWs and CTQWs have been established through reduction of the DTQW's larger state space by taking appropriate limits \cite{PhysRevA.74.030301,DAlessandro201085,Childs2009,MD12,PhysRevA.91.062304}. Strauch~\cite{PhysRevA.74.030301} analyzed the connection between coined and continuous-time QWs on the line by describing a method to convert the evolution equations of the coined model into the evolution equations of the CTQW model when the length of the time steps tends to zero.
D'Alessandro~\cite{DAlessandro201085} extended Strauch's results by first obtaining the dynamics of CTQWs on $d$-dimensional lattices as an appropriate limit of the dynamics of the coined model on the same lattices, and by then extending those results to regular graphs. Childs~\cite{Childs2009} used the Szegedy QW model~\cite{1366222} to propose a coined model whose behavior approaches that of a related CTQW in a certain limit. Molfetta and Debbasch~\cite{MD12} analyzed the same kind of connection in the case when both time and length steps tend to zero. For DTQWs on $d$-regular graphs with coins that are both unitary and Hermitian, Dheeraj and Brun~\cite{PhysRevA.91.062304} presented constructions of families of DTQWs that have well defined continuous-time limits on larger graphs. In this work, we go in the opposite direction; instead of obtaining CTQWs as limits of DTQWs, we produce the evolution of the coined discrete-time model on a generic graph using a CTQW method. 

Percolation graphs and statistical networks introduce decoherence into QWs, causing a shift towards classical behavior. This was noted for the first time by Romanelli et al.~\cite{Romanelli:2005} for coined QWs on the line. It was generalized to 2-dimensional lattices in Ref.~\cite{Oliveira:2006} and analyzed further in Ref.~\cite{KKNJ12}. Continuous-time QWs on percolation graphs were addressed in many papers, such as Refs.~\cite{PhysRevE.76.051125,Anishchenko2012,1751-8121-46-37-375305}. In the usual percolation model, the graph changes dynamically as edges break randomly or are inserted randomly. The ``percolation'' that is used in our construction is different in that it is systematic rather than random, and it so creates a graph alternation that allows to define CTQWs that are equivalent to given coined QWs. Our work borrows some ideas from the staggered model \cite{Portugal2016staggered}, which can be considered an intermediate step for our constructions and which was also used in Ref.~\cite{Portugal2016} to connect Szegedy's model with coined QWs. It is not necessary to be familiar with the staggered model though, as the paper at hand is entirely self-contained. 

In this work we will, starting from the coined model on a generic graph, define a CTQW on a larger graph, which we call the expanded graph, that exactly reproduces the evolution of the original coined QW. Instances of this expanded graph have already appeared in Ref.~\cite{PhysRevA.91.062304}, which has similarities to our work but a different objective. We consider flip-flop coined QWs on undirected graphs. This type of DTQW acts on Hilbert spaces of dimension $2\abs{E}$, where $\abs{E}$ is the number of edges. Since the dimension of the Hilbert space in which a CTQW takes place is equal to the size of the graph, the number of nodes of the expanded graph has to be $2\abs{E}$. The most simple case is when the graph is regular and when only Grover coins are used, and then we will obtain correspondence of the coined QW to continuous-time evolution of the form $e^{-itH}$, where the Hamiltonian $H$ is the graph Laplacian and the underlying graph changes dynamically between two percolations of the expanded graph. It would be desirable to avoid the use of percolation; however, since our constructions are exact, they thereby provide insights into the limitations for attempts to reconcile the coined and continuous-time models. We hence consider our work a new starting point for further studies of that relation.

This paper is structured as follows. First we review CTQWs and combine them with percolation \cite{PhysRevLett.49.486,Santos2014} (Sec.~\ref{sec:CTQW}). We then review flip-flop coined QWs (Sec.~\ref{sec:DTQW}), a commonly used type of DTQWs, and take first steps to translate their building blocks into the continuous-time setting (Sec.~\ref{sec:ops_on_cliques}). After that, we present our construction of simulations of flip-flop coined QWs by means of CTQWs on larger, percolated graphs (Sec.~\ref{sec:construction}). We then carry out that construction for a simple example (Sec.~\ref{sec:example}) and conclude our work (Sec.~\ref{sec:conclusion}).

\section{Percolated continuous-time QWs}
\label{sec:CTQW}

\subsection{Standard continuous-time QWs}

We first briefly review the notion of a continuous-time quantum walk (CTQW) on an undirected graph $G(V,E)$. The state space $\mathcal{H}$ for a CTQW on $G$ has dimension $\abs{V}$, and we use the set of vertices $V$ as the computational basis,
\begin{equation*}
 \mathcal{H} = \text{span} \{ \ket{0}, \ket{1}, \dots, \ket{\abs{V}-1}\}
.\end{equation*}
A particle in the graph is described by a state $\ket{\psi}\in\mathcal{H}$, and the quantity
\begin{equation*}
 p_k = \abs{\langle k | \psi \rangle}^2
\end{equation*}
is the probability that it is found at vertex $k$. A \emph{continuous-time quantum walk} is the time evolution an initial state $\ket{\psi_0}$ undergoes through the action of a propagator $U(t)=e^{-itH}$, that is
\begin{equation*}
 \ket{\psi(t)} = e^{-itH}\ket{\psi_0},
\end{equation*}
where $H$ is a Hermitian operator on $\mathcal{H}$.

The operator $H$ is the Hamiltonian of the system and in the context of CTQWs on graphs, the \emph{graph Laplacian} $L_{\text{g}}=D-A$ is a common choice for it. Here, $A$ is the adjacency matrix of the graph and the degree matrix $D$ is diagonal with entries $d_{ii} = \sum_k a_{ik}$, where $a_{ik}$ is the $(i,k)$-th entry of $A$. We also apply this definition to weighted graphs; the degree of a vertex is the sum of the weights of all incident edges. 

All graphs in this paper are assumed to be free of loops, i.e. the diagonal entries of the adjacency matrix are assumed to be zero. While loops have no effect on the graph Laplacian, their presence would unnecessarily complicate some of our later constructions.

\subsection{Continuous-time QWs with percolation}

We now introduce \emph{percolation} into the CTQW model. In percolation theory, which is, for example, used to study the flow of liquids in porous media, edges in a graph are randomly broken or inserted with some fixed probability. In order to establish the connection between coined and continuous-time QW models, we need to be able to switch on and off edges as well. However, we will be doing so in a very systematic way, which does not generate decoherence.

As an example for percolation, consider a line segment $G(\{1,2,3\},\{\{1,2\},\{2,3\}\})$ with $3$ vertices, where edge $\{1,2\}$ has weight $1$ and edge $\{2,3\}$ has weight $2$. The below expression shows how the graph Laplacian is built, and how it changes when the edge with weight $2$ is switched off. In particular, the percolation affects not only the two entries of $L_{\text{g}}$ that correspond to the edge that is being broken or inserted, but it also adjusts the diagonal entries of the involved vertices, i.e. their degrees, accordingly:
\begin{equation*}
 A=\begin{bmatrix*}[r] 0 & 1 & 0 \\ 1 & 0 & 2 \\ 0 & 2 & 0 \end{bmatrix*} \quad \leadsto
 L_{\text{g}}=\begin{bmatrix*}[r] 1 & -1 & 0\phantom{-} \\ -1 & 3 & -2\phantom{-} \\ 0 & -2 & 2\phantom{-} \end{bmatrix*} \quad \leadsto
 L_{\text{g}}'=\begin{bmatrix*}[r] 1 & -1 & 0\phantom{-} \\ -1 & 1 & 0\phantom{-} \\ 0 & 0 & \phantom{-}0\phantom{-} \end{bmatrix*} 
.\end{equation*}

In the CTQW we will derive later, the Hamiltonian is either the graph Laplacian or a simple derivation thereof. During the continuous-time evolution we will then switch on and off certain edges of the graph. This allows to accommodate the alternation of coin steps and shift steps in the discrete-time model with \emph{one} Hamiltonian in the continuous-time model, that changes only implicitly via percolation of the underlying graph.

We conclude our discussion of the continuous-time setting by making the following observation on graph Laplacians of percolated graphs. If we derive the graphs $G_1(V,E_1)$ and $G_2(V,E_2)$ from $G(V,E)$ through percolation such that $E_1 \cap E_2 = \emptyset, E_1 \cup E_2 = E$, then 
\begin{equation*}
L_{\text{g}}(G_1)+L_{\text{g}}(G_2)=L_{\text{g}}(G)
.\end{equation*}

\section{Coined QWs}
\label{sec:DTQW}

\subsection{Flip-flop coined QWs}

We now describe flip-flop coined QWs on undirected graphs $G(V,E)$, where $V$ is the set of vertices and $E$ is the set of edges. Let the vertices be labeled by $0,1,2,\dots,\abs{V}-1$ and the edges by $0,1,2,\dots,\abs{E}-1$. Use $v$ to denote a generic vertex and $j$ for a generic edge. We define the set of vertex-edge pairs
\begin{equation*}
 \Gamma = \{ (v,j) \: | \: v \in V, j \in E, v \text{ is an endpoint of } j \}
\end{equation*}
and further the subsets
\begin{equation*}\begin{split}
 \Gamma^{v} & = \{ (v,j') \: | \: j' \in E \text{ with } (v,j') \in \Gamma \}, \\
 \Gamma_{j} & = \{ (v',j) \: | \: v' \in V \text{ with } (v',j) \in \Gamma \}
\end{split}\end{equation*}
of $\Gamma$. We have $\abs{{\Gamma_j}}=2$ for all $j$, and $\abs{\Gamma^v}$ is equal to the degree of the vertex $v$.

For a flip-flop coined QW on $G$, we consider the $(2\abs{E})$-dimensional Hilbert space $\mathcal{H}$ that is spanned by the states
\begin{equation}\label{eq:comp_basis}
 \{ \ket{v,j} \: | \: (v,j) \in \Gamma \}
.\end{equation}
For notational simplicity, we denote the span of this set again by $\Gamma$ (i.e. $\mathcal{H}=\Gamma$) and we define the subspaces $\Gamma_j$ and $\Gamma^v$ accordingly. For $\Gamma$ itself as well as for the subspaces $\Gamma_j$ and $\Gamma^v$, we always use \eqref{eq:comp_basis} or subsets thereof as the computational basis. The \emph{flip-flop coined quantum walk} is driven by the operator
\begin{equation*}
 U = S \circ C
,\end{equation*}
where the coin $C$ is a direct sum of unitary operators acting on the spaces $\Gamma^v$. An additional restriction on the form of $C$, that is specific to the notes at hand, is stated in the next section, cf.~\eqref{eq:form_of_Cv_several}. The shift $S$ is defined by
\begin{equation}\label{eq:action_of_S}
 S\ket{v,j} = \ket{v',j}
,\end{equation}
where $\Gamma_j=\{(v,j),(v',j)\}$.
Note that $S$ is a direct sum\footnote{In order to write down a matrix representation of $C$ or $S$, one first has to decide how to list the basis elements $\ket{v,j}$. Arranging them with respect to $v$ yields a matrix representation of $C$ that is in block diagonal form. The representation of $S$ can be made block diagonal by ordering in the $j$-component.} of operators acting on the spaces $\Gamma_j$, and that $S^2=I$. 

The flip-flop coined QW takes its name from definition~\eqref{eq:action_of_S}, which states that, during a shift step, a quantum walker standing at vertex $v$ and facing edge $j$ walks along this edge and then turns around to again face edge $j$. The fact that this definition does not depend on the structure of the graph or on the labeling of its vertices and edges is a strong argument for the significance of flip-flop coined QWs.

Restricting our attention to one of the subspaces $\Gamma_j$, we find that the matrix representation of $S$ is rather simple:
\begin{equation*}
 S_{|\Gamma_j} = S_j = \begin{bmatrix} 0 & 1 \\ 1 & 0 \end{bmatrix}
.\end{equation*}
Letting $\ket{s}$ be the normalized uniform distribution on the $2$-dimensional space $\Gamma_j$, we can also write
\begin{equation}\label{eq:form_of_Sj}
 S_j = 2\ket{s}\bra{s} - I
.\end{equation}

\subsection{Admissible coins and the expanded graph}

The additional requirement for $C$ is that all its components are of a form similar to \eqref{eq:form_of_Sj}, namely
\begin{equation}\label{eq:form_of_Cv_several}
 C_{|\Gamma^v} = C^v = 2 \sum_{k} \ket{\alpha_k}\bra{\alpha_k} - I
,\end{equation}
where $\{\ket{\alpha_k}\}$ is an orthonormal set in $\Gamma^v$. Such coins are, in addition to being unitary, Hermitian, and we have $(C^v)^2 = I$, which is why they are called reflections. However, we first present our construction for coins of the form
\begin{equation}\label{eq:form_of_Cv_one}
 C^v = 2 \ket{\alpha}\bra{\alpha} - I
,\end{equation}
where $\ket{\alpha}$ is a normalized state of $\Gamma^v$, and we will then outline the case~\eqref{eq:form_of_Cv_several} at the end of our analysis (in Sec~\ref{sec:general_coins}, which also provides an overview of the coins covered in this work). Note that choosing $\ket{\alpha}=\ket{s}$ gives the Grover coin. We allow the set $\{\ket{\alpha_k}\}$ in~\eqref{eq:form_of_Cv_several} to be empty. In this case we get $C^v=-I$, which we call the search coin.  

\begin{figure}
\begin{center}
\begin{tabular}{ccc}
\raisebox{0.4\height}{\includegraphics[scale=.65]{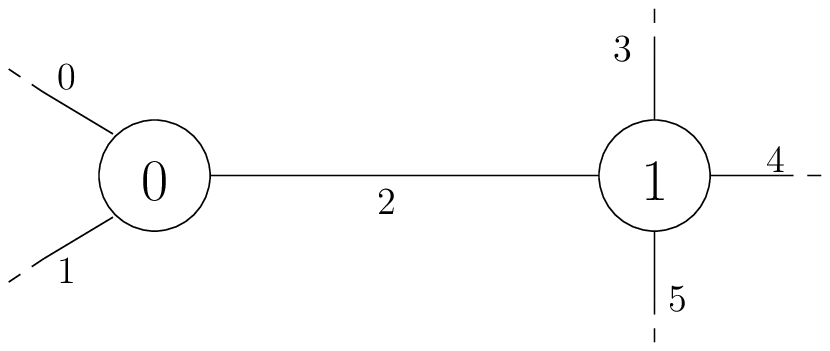}}
& \phantom{abcdefgh} & 
\includegraphics[scale=.65]{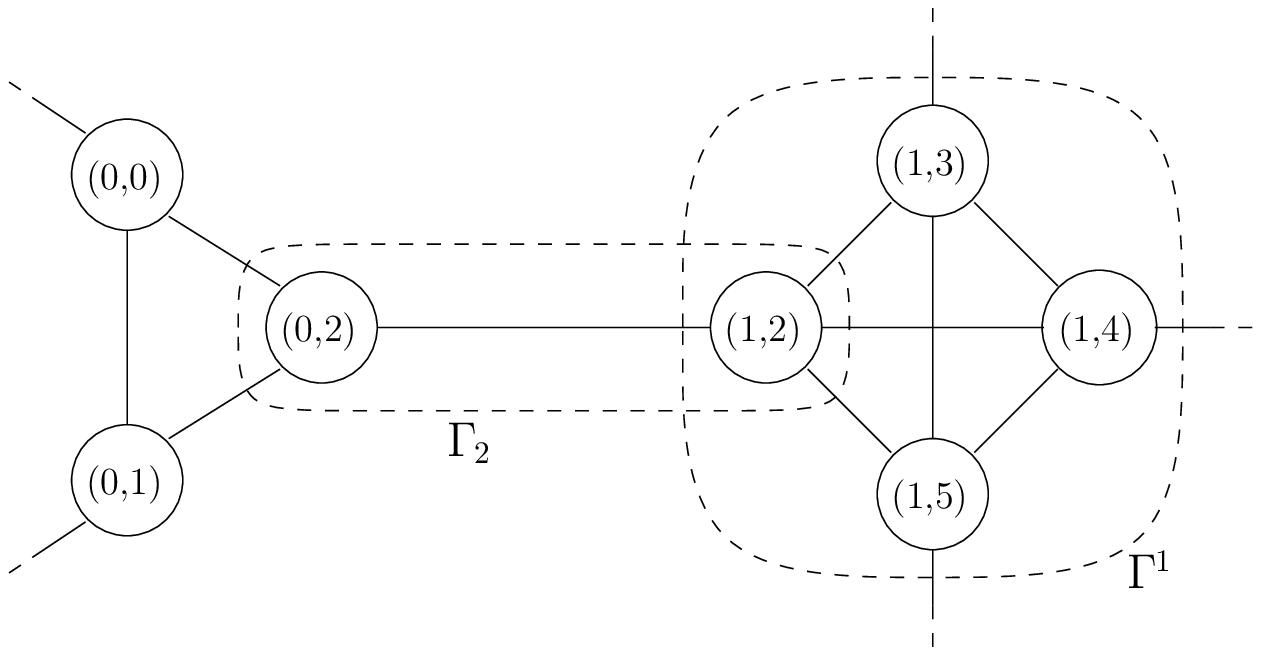}
\end{tabular}
\end{center}
\caption[Expansion]{Two vertices of degrees $3$ and $4$ in the original graph $G$ (left) are expanded to cliques of sizes $3$ and $4$ in the expanded graph $G_{\text{exp}}$ (right).}\label{fig:graphs_expansion}
\end{figure}

We now present a visualization of the set of basis elements~\eqref{eq:comp_basis} of $\Gamma$ that preserves the global structure of the graph $G$. Consider a vertex $v$ of degree $d$ with incident edges $\{j_0,j_1,\dots,j_{d-1}\}$. We replace $v$ by a clique of size $d$ and label its vertices $(v,j_0),(v,j_1),\dots,(v,j_{d-1})$. Each vertex $(v,j_k)$ is connected to exactly one other clique, namely the one that replaced the vertex the original edge $j_k$ lead to. Let us call this larger graph $G_{\text{exp}}$, the \emph{expanded graph}. Figure~\ref{fig:graphs_expansion} illustrates the expansion $G \leadsto G_{\text{exp}}$ and also the subsets of vertices that span the subspaces $\Gamma_j$ and $\Gamma^v$, which will be of great importance in later constructions. Note that the Hilbert space for a CTQW on $G_{\text{exp}}$ has the same dimension as $\Gamma$, the Hilbert space for a coined QW on $G$.

In the next section, we focus on one building block $C^v$ or $S_j$ of the discrete-time propagator $U$, and we find Hamiltonians whose continuous-time propagators produce the same evolution over certain time intervals. That theory will later be applied to the subspaces $\Gamma_j$ and $\Gamma^v$. Hence the basis below can be thought of as the corresponding subset of~\eqref{eq:comp_basis}, and its size $N$ is either equal to $2$ or to the degree of some vertex in the graph $G$.

\section{Operators on cliques}
\label{sec:ops_on_cliques}

Consider an $N$-dimensional space with basis $\{\ket{0},\ket{1},\dots,\ket{N-1}\}$ and define the operators
\begin{equation*}
 A_{\ket{\alpha}} = 2\ket{\alpha}\bra{\alpha} - I
,\end{equation*}
where $\ket{\alpha}$ is of unit norm. Further define the operator $L_{\ket{\alpha}}$ via
\begin{equation}\label{eq:AandL}
 A_{\ket{\alpha}} = e^{-i \tau L_{\ket{\alpha}}}
.\end{equation}
We will specify how this is made well defined and we also determine the parameter $\tau$. Our goal in this section is to find $L_{\ket{\alpha}}$ for certain choices of $\ket{\alpha}$, and to see how it transforms when $\ket{\alpha}$ is changed.

We have $A_{\ket{\alpha}}\ket{\alpha} = \ket{\alpha}$ and $A_{\ket{\alpha}}\ket{\beta} = - \ket{\beta}$ for any $\ket{\beta}$ that is orthogonal to $\ket{\alpha}$. Hence, by extending $\{\ket{\alpha}\}$ to an orthonormal basis, we find the spectrum
\begin{equation*}
\sigma(A_{\ket{\alpha}}) = \sigma(e^{-i \tau L_{\ket{\alpha}}}) = \{ +1, -1, -1, \dots, -1\}
.\end{equation*}
We would like $L_{\ket{\alpha}}$ to be of the  form of a Laplacian---at least for some choices of $\ket{\alpha}$---and therefore we choose the branch of the logarithm so that the single eigenvalue of $L_{\ket{\alpha}}$ is $0$ and the others are positive. We further set $\tau = \frac{\pi}{N}$ to obtain the spectrum of the graph Laplacian on a complete graph:
\begin{equation*}
 \sigma(L_{\ket{\alpha}}) =  \Big\{ 0, N, N, \dots, N \Big\} 
.\end{equation*}
 
It is now important to understand how $L_{\ket{\alpha}}$ changes if $\ket{\alpha}$ in \eqref{eq:AandL} is replaced by some other unit state $\ket{\alpha'}$. There exists a unitary transformation $W$ with $\ket{\alpha'} = W \ket{\alpha}$. This gives
\begin{equation*}
 A_{\ket{\alpha'}} = WA_{\ket{\alpha}}W^\dagger = e^{-i\frac{\pi}{N}WL_{\ket{\alpha}}W^\dagger}
,\end{equation*}
and we see that upon changing $\ket{\alpha}$, the operator $L_{\ket{\alpha}}$ transforms as follows:
\begin{equation}\label{eq:trafo_L}
 L_{\ket{\alpha}} \leadsto W L_{\ket{\alpha}} W^\dagger
.\end{equation}
For $\ket{\alpha} = \ket{0}$, we have $A_{\ket{0}} = \text{diag}(+1,-1,-1,\dots,-1)$, and we find
\begin{equation}\label{eq:1cgL}
 L_{\ket{0}} = \text{diag}(0,N,N,\dots,N)
.\end{equation}
More generally, for $\ket{\alpha}=\ket{k}$ with $ 0 \le k < N $, the position of the zero entry on the diagonal is the $(k+1)$-th place.

To find $L_{\ket{s}}$, we use vector notation and define the matrix
\begin{equation*}
 W = \begin{bmatrix}
      1/\sqrt{N} & | & | & & | \\
      1/\sqrt{N} & b_2 & b_3 & \dots & b_N \\
      1/\sqrt{N} & | & | & & | \\
      \vdots & \vdots & \vdots & & \vdots \\
      1/\sqrt{N} & | & | & & | \\
     \end{bmatrix}
  = \begin{bmatrix}
     s & b_2 & b_3 & \dots & b_N
    \end{bmatrix}
,\end{equation*}
where the $b_k$ are arbitrary vectors that extend $\{s\}$ to an orthonormal basis. Then we have $\ket{s} = W\ket{0}$, $L_{\ket{0}}=N\,(I-\ket{0}\bra{0})$, and we see that $L_{\ket{s}}$ is the graph Laplacian of the complete graph:
\begin{equation}\label{eq:scgL}
 L_{\ket{s}} = WL_{\ket{0}}W^\dagger 
 = NI-N s s^\dagger = \begin{bmatrix}
                N-1 & -1 & -1 & \dots & -1 \\
                -1 & N-1 & -1 & \dots & -1 \\
                -1 & -1 & N-1 &  & -1 \\
                \vdots & \vdots &  & \ddots & \vdots \\
                -1 & -1 & -1 & \dots & N-1
               \end{bmatrix}
.\end{equation}
Note that this computation also shows that the nonuniqueness of the transformation $W$ is not problematic.

\section{Simulating coined QWs with CTQWs}
\label{sec:construction}

\begin{figure}
\begin{center}
\includegraphics[scale=.35]{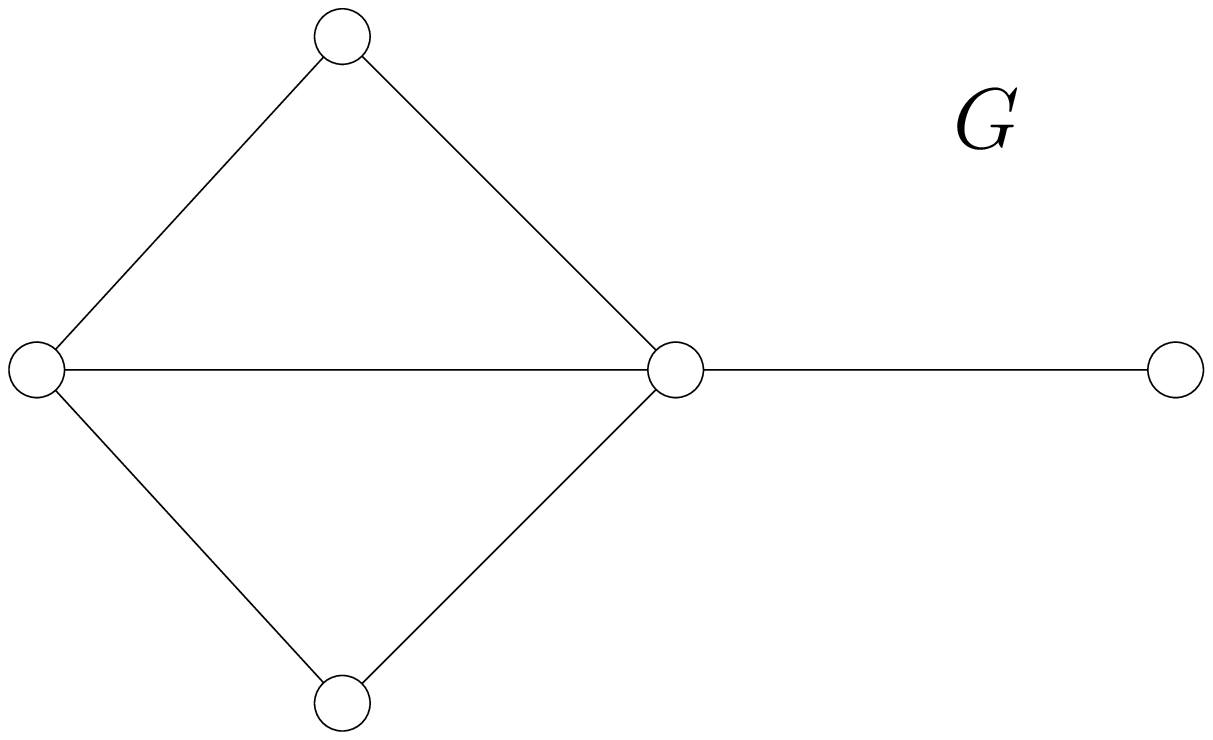} \\ ~ \\ ~ \\
\includegraphics[scale=.35]{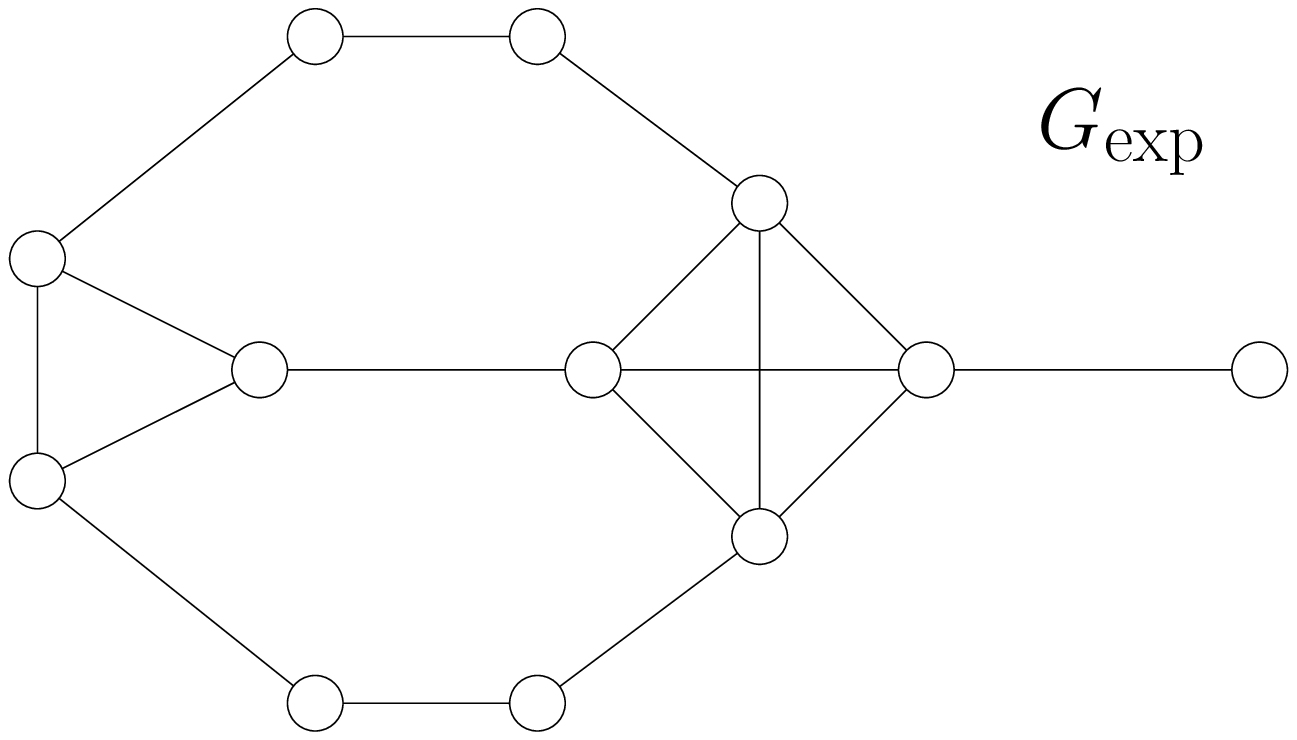} \\ ~ \\
\begin{tabular}{cc}
\includegraphics[scale=.35]{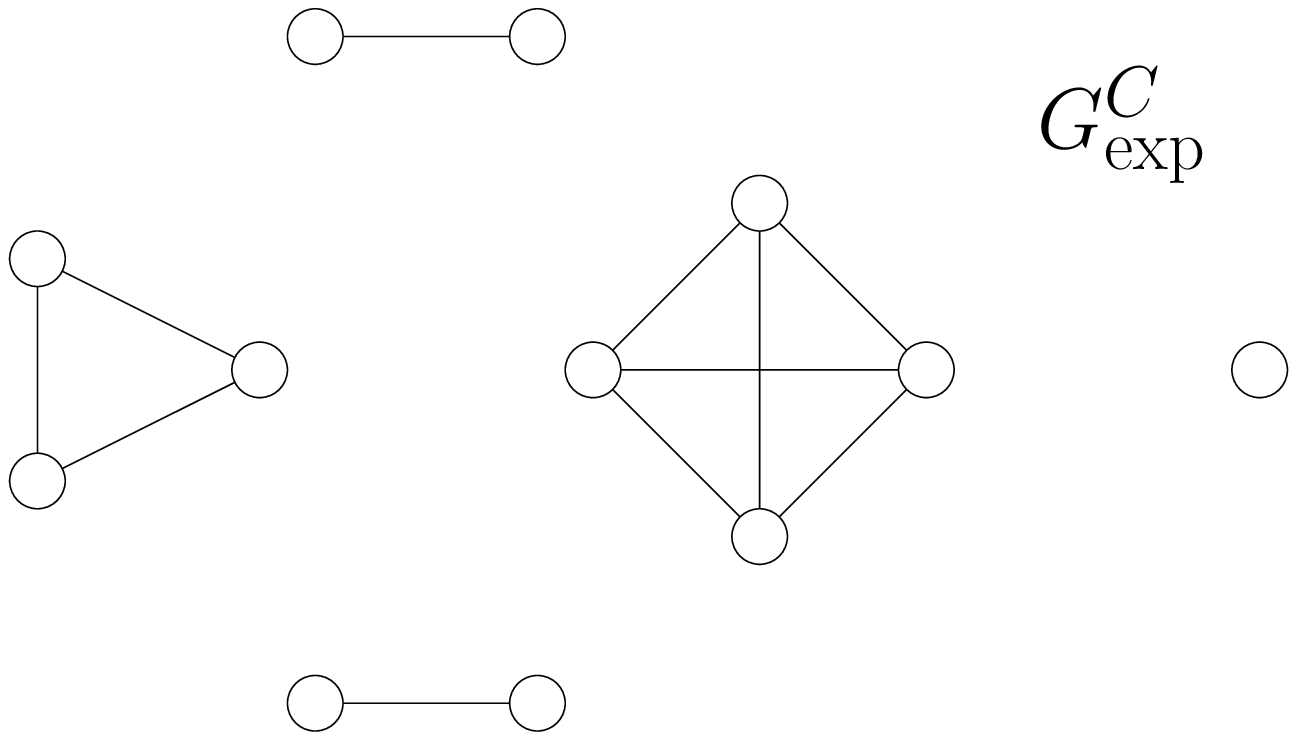} \phantom{spacespace} & \phantom{spacespace}
\includegraphics[scale=.35]{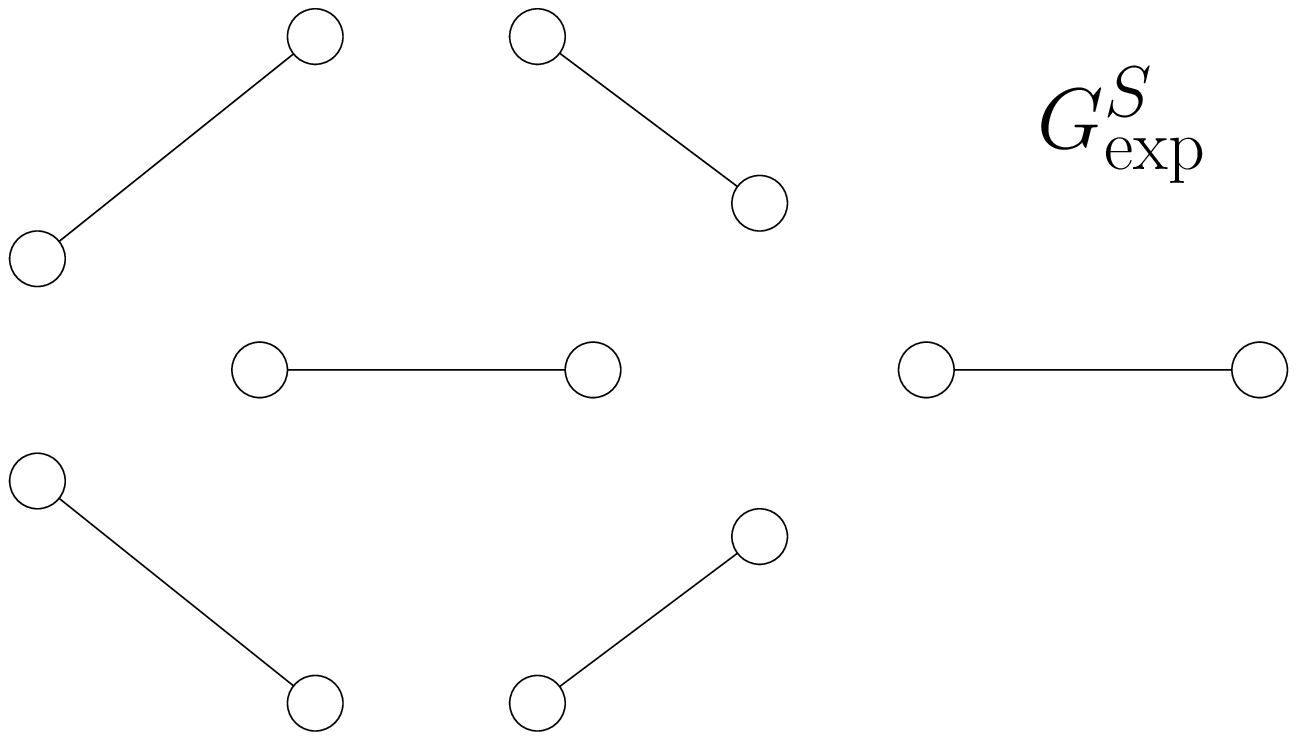}
\end{tabular}
\end{center}
\caption[Percolation]{The graph $G(V,E)$ is expanded to the graph $G_{\text{exp}}$ with $2\abs{E}$ vertices. From $G_{\text{exp}}$, the graphs $G_{\text{exp}}^C$ and $G_{\text{exp}}^S$ are derived via percolation. The cliques of $G_{\text{exp}}^C$ correspond to vertices of the original graph $G$, while the 2-cliques of $G_{\text{exp}}^S$ correspond to edges of $G$.}\label{fig:graph_percolation}
\end{figure}

We now construct CTQWs on $G_{\text{exp}}$ that reproduce the evolution of flip-flop coined QWs on $G$ satisfying~\eqref{eq:form_of_Cv_one}. In order to achieve this, we need to introduce percolation, i.e. allow edges of $G_{\text{exp}}$ to be switched on and off during the continuous-time evolution. As mentioned earlier, this will be done in a very systematic way (rather than randomly, which is the modus operandi for percolation theory; cf. Sec.~\ref{sec:example}). Let $G_{\text{exp}}^{C}$ be the graph we obtain from $G_{\text{exp}}$ by removing the $\abs{E}$ edges that lie within $\Gamma_j$ components (in the expanded graph on the right-hand side of Fig.~\ref{fig:graphs_expansion}, those are the edge between $(0,2)$ and $(1,2)$ and the five edges whose second endpoints are not included in the figure). We will see that this subgraph, which consists of $\abs{V}$ separated cliques, is the supporting structure for the evolution perpetuated by the coin step $C$ in the discrete-time model. We derive a second graph, $G_{\text{exp}}^{S}$, by removing all edges that lie within $\Gamma^v$ components (in Fig.~\ref{fig:graphs_expansion}, the three edges that connect vertices $(0,\cdot)$ and the six edges between vertices $(1,\cdot)$). This subgraph consists of $\abs{E}$ separated cliques of size $2$. An illustration of this construction is given by Fig.~\ref{fig:graph_percolation}.

\subsection{QWs with Grover coins}

Recall that the components of the shift operator $S$ act on 2-cliques and are given by $S_j=A_{\ket{s}}$, cf.~\eqref{eq:form_of_Sj}. By~\eqref{eq:scgL}, the corresponding continuous-time Hamiltonian $L_{\ket{s}}$ that produces the same evolution over a time interval of length $\tau = \frac{\pi}{2}$ is the graph Laplacian of that 2-clique. The direct sum of these operators $L_{\ket{s}}$ on the different components $\Gamma_j$ is the graph Laplacian on $G_{\text{exp}}^{S}$. Now, if Grover coins are used at all vertices of $G$, then we have $C^v = A_{\ket{s}}$ for all $v$ and consequently continuous-time evolution with respect to the graph Laplacian in all components of $G_{\text{exp}}^{C}$ as well. This leads to the following observation. If the graph $G$ is $d$-regular and if Grover coins are used at all vertices, then the evolution for one time step in the flip-flop coined QW on $G$ with propagator $ U = S \circ C $ is reproduced by the following percolated CTQW on the expanded graph $G_{\text{exp}}$: 
\begin{enumerate}
\item[(a)]  Evolve the system with respect to the graph Laplacian on $G_{\text{exp}}^{C}$ for time $\frac{\pi}{d}$, and then
\item[(b)] evolve with respect to the graph Laplacian on $G_{\text{exp}}^{S}$ for time $\frac{\pi}{2}$.
\end{enumerate}
That is, the Hamiltonian of the system is the graph Laplacian of $G_{\text{exp}}$; it changes only implicitly due to percolation of $G_{\text{exp}}$ at times $t=0$, $t=\frac{\pi}{d}$, $t=\frac{\pi}{d}+\frac{\pi}{2}$, $t=\frac{2\pi}{d}+\frac{\pi}{2}$, etc., and it so takes the two forms needed for the coin step~(a) and the shift step~(b). Hence we have established a correspondence to propagation of the form $e^{-itH}$ with \emph{one} Hamiltonian $H$, i.e., to a CTQW. Steps (a) and (b) correspond to the lines 7--9 and 10--12, respectively, of the algorithm in Sec.~\ref{sec:example} (with a small modification in the evolution time for (a)).

If the graph $G$ is not $d$-regular, we simply weight the components of the graph Laplacian $L_{\text{g}}$ of $G_{\text{exp}}^C$. To be more precise, let
\begin{equation}\label{eq:not_regular}
 L^C = \bigoplus_{v \in V} \frac{2}{d_v} L_{\text{g}}(G_{\text{exp}}^C)_{|\Gamma^v} 
,\end{equation}
where $d_v$ is the degree of the vertex $v$, i.e. the dimension of $\Gamma^v$. Using the operator $L_{\text{g}}(G_{\text{exp}}^S) + L^C$ as the Hamiltonian on the graph $G_{\text{exp}}$, we now obtain equivalence of the two types of QWs as above, but with percolations after every $\frac{\pi}{2}$ time units (cf.~\eqref{eq:AandL} and recall that $\tau=\frac{\pi}{N}$).

\subsection{QWs with general coins}
\label{sec:general_coins}

We now consider the situation when coins of the form~\eqref{eq:form_of_Cv_one} with $\ket{\alpha}\not=\ket{s}$ are used. There are normalized states $\ket{\alpha^{0}}\in\Gamma^0,\ket{\alpha^{1}}\in\Gamma^1,\dots,\ket{\alpha^{\abs{V}-1}}\in\Gamma^{\abs{V}-1}$ such that
\begin{equation*}
 C = \bigoplus_{v \in V} A_{\ket{\alpha^{v}}} = \bigoplus_{v \in V} 2\ket{\alpha^{v}}\bra{\alpha^{v}} - I
.\end{equation*}
Let $W_{0}, W_{1}, \dots,W_{\abs{V}-1}$ be a collection of unitary transformations such that $W_{v}$ maps the uniform distribution $\ket{s^{v}}$ of size $\abs{\Gamma^v}$ to $\ket{\alpha^{v}}$. Defining the operator
\begin{equation*}
 W = \bigoplus_{v \in V} W_{v}
\end{equation*}
allows to carry out the transformations~\eqref{eq:trafo_L} in all subspaces $\Gamma^v$ simultaneously. Hence we obtain the above correspondence between the flip-flop coined QW on $G$ and the CTQW on $G_{\text{exp}}$ after transforming the operator $L^C$ in~\eqref{eq:not_regular} with $W$, that is 
\begin{equation*}
 L^C \leadsto W L^C W^\dagger
.\end{equation*}

If marked vertices, at which search coins $-I$ are used, are present, then we replace the corresponding components of $L^C$ in~\eqref{eq:not_regular} by $2I$ (cf.~\eqref{eq:AandL} with $\tau=\frac{\pi}{2}$).

We now briefly sketch the construction of the continuous-time Hamiltonian for coins of the form
\begin{equation}\label{eq:form_of_Cv_several_2}
 C^v = 2 \sum_{k=0}^{m-1} \ket{\alpha_k}\bra{\alpha_k} - I
.\end{equation}
Here the vertex $v$ of the original graph is fixed, $\{\ket{\alpha_k}\}$ is an orthonormal set in $\Gamma^v$, and $m$ is less than or equal to the degree $d$ of $v$. Define the transformation
\begin{equation*}
 \widetilde{W} = \ket{\alpha_0}\bra{0} + \ket{\alpha_1}\bra{1} + \dots + \ket{\alpha_{m-1}}\bra{m-1} + \ket{\beta_m}\bra{m} + \dots + \ket{\beta_{d-1}}\bra{d-1}
,\end{equation*}
where $\{\ket{k}\}$ is the standard basis of $\Gamma^v$ (i.e. a subset of~\eqref{eq:comp_basis}) and the $\ket{\beta_k}$ are arbitrary states that extend $\{\ket{\alpha_k}\}$ to an orthonormal basis of $\Gamma^v$. Then we have
\begin{equation*}
 \widetilde{C} = \widetilde{W}^{\dagger}C^v\widetilde{W} = \text{diag}(+1,\dots,+1,-1,\dots,-1)
,\end{equation*}
for which $\widetilde{C}=e^{-i \tau \widetilde{L}}$ is easily solved, cf.~\eqref{eq:1cgL}. Applying $\widetilde{W}$ from the left and $\widetilde{W}^{\dagger}$ from the right then yields a Hamiltonian whose propagator over the time interval $\tau$ agrees with the discrete-time coin step $C^v$.

We conclude the theoretical part of these notes by summarizing the scope of our work. For a flip-flop coined QW on a graph $G$ that uses only coins of the form~\eqref{eq:form_of_Cv_several_2}, we have produced an equivalent CTQW on a larger, dynamically percolated graph. The graph $G$ need not be regular, and coins can differ between different vertices. For example, our theory covers search coins $C_S$ ($m=0$, where $m$ is the number of terms in the sum~\eqref{eq:form_of_Cv_several_2}), Grover coins $C_G$ ($m=1$, $\ket{\alpha}=\ket{s}$), the $2$-dimensional Hadamard coin $C_{H,2}$ ($N=d=2$, $m=1$, $\ket{\alpha}=\frac{(2+\sqrt{2})\ket{0}+\sqrt{2}\ket{1}}{2\sqrt{2+\sqrt{2}}}$), and the $4$-dimensional Hadamard coin $C_{H,4} = C_{H,2} \otimes C_{H,2}$ ($N=d=4$, $m=2$, $\ket{\alpha_0}=\frac{\ket{1}+\ket{2}}{\sqrt{6}}-\sqrt{\frac23}\,\ket{3}$, $\ket{\alpha_1}=\frac{\sqrt{3}}{2}\,\ket{0}+\frac{\ket{1}+\ket{2}+\ket{3}}{2\sqrt{3}}$). However, there are coins to which the constructions in this paper do not apply, namely coins that are not reflections. An example is given by the $4$-dimensional Fourier coin $C_{F,4}$, which has a complex eigenvalue while for the reflections $C^v$ in~\eqref{eq:form_of_Cv_several_2} we have $\sigma(C^v)\subseteq\{+1,-1\}$.

\section{Example}
\label{sec:example}

\begin{figure}
\begin{center}
\includegraphics[scale=.7]{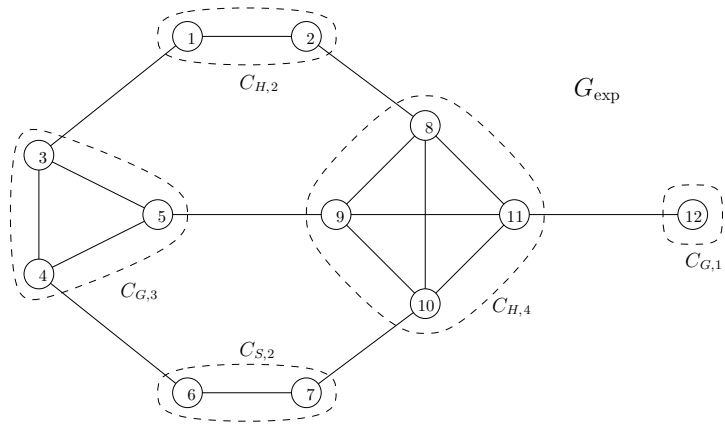}
\end{center}
\caption[Example]{The labeling of the expanded graph for the sample construction in Sec.~\ref{sec:example}. At the vertices of the original graph $G$ from Fig.~\ref{fig:graph_percolation}, which are represented by the cliques encircled by dashed curves, different coins are used: Grover coins ($C_G$; the second subindex denotes the dimension of the respective coin space), Hadamard coins ($C_H$), and a search coin ($C_S$).}\label{fig:graph_example}
\end{figure}

As an example for our construction, we now state the continuous-time Hamiltonian that reproduces the evolution of a flip-flop coined QW on the graph $G$ in Fig.~\ref{fig:graph_percolation}. Figure~\ref{fig:graph_example} shows which coins are used and, for the matrix representation~\eqref{eq:hamil_example}, how the vertices of the expanded graph are labeled. The Hamiltonian $H$ on $G_{\text{exp}}$ is
\begin{equation}\label{eq:hamil_example}
H= \begin{bsmallmatrix*}[c]
  \frac{1}{2+\sqrt{2}} +1 & -\frac{1}{\sqrt{2}} & -1 & & & & & & & & & \\[0.3em]
  -\frac{1}{\sqrt{2}} & \frac{2+\sqrt{2}}{2} +1 & & & & & & -1 & & & & \\[0.3em]
  -1 & & \frac{4}{3} +1 & -\frac{2}{3} & -\frac{2}{3} & & & & & & & \\[0.3em]
  & & -\frac{2}{3} & \frac{4}{3} +1 & -\frac{2}{3} & -1 & & & & & & \\[0.3em]
  & & -\frac{2}{3} & -\frac{2}{3} & \frac{4}{3} +1 & & & & -1 & & & \\[0.3em]
  & & & -1 & & 2+1 & 0 & & & & & \\[0.3em]
  & & & & & 0 & 2+1 & & & -1 & & \\[0.3em]
  & -1 & & & & & & \frac12 +1 & -\frac12 & -\frac12 & -\frac12 & \\[0.3em]
  & & & & -1 & & & -\frac12 & \frac32 +1 & -\frac12 & \frac12 & \\[0.3em]
  & & & & & & -1 & -\frac12 & -\frac12 & \frac32 +1 & \frac12 & \\[0.3em]
  & & & & & & & -\frac12 & \frac12 & \frac12 & \frac12 +1 & -1 \\[0.3em]
  & & & & & & & & & & -1 & 0 + 1
 \end{bsmallmatrix*}
.\end{equation}
The $-1$ entries and the $+1$ summands on the diagonal of~\eqref{eq:hamil_example} come from the graph Laplacian on $G_{\text{exp}}^S$. Removing them from the matrix, which is effectuated by the percolation step in line 7 of the algorithm below, we are left with $L^C$, the operator on $G_{\text{exp}}^C$ whose continuous-time evolution over a time interval of length $\frac\pi2$ agrees with the original coin operator $C$. Due to our labeling of the vertices of $G_{\text{exp}}$, cf. Fig.~\ref{fig:graph_example}, the matrix $L^C$ is block diagonal. Its second block, for example, corresponds to a Grover coin and hence is the graph Laplacian of a complete graph, weighted according to~\eqref{eq:not_regular}.

We now state the general algorithm for continuous-time simulation of flip-flop coined QWs. For the reference to ``dashed curves'' in these instructions, compare to Fig.~\ref{fig:graph_example}.

\begin{algorithm}[H]
\caption{Continuous-time simulation of a flip-flop coined QW}
\begin{algorithmic}[1]
	\REQUIRE graph $G$ with specification of coins, initial state $\ket{\psi_0}$, time $t_f$
	\ENSURE the final state $\ket{\psi_{t_f}}$ of the flip-flop coined QW on $G$ after $t_f$ steps
	\STATE $t\gets0$
	\STATE construct the expanded graph $G_{\text{exp}}$ described in Sec.~\ref{sec:DTQW} (e.g. Fig.~\ref{fig:graph_example})
	\STATE set $\ket{\psi_0}$ as the initial state for the CTQW on $G_{\text{exp}}$ \\ (the vertices of $G_{\text{exp}}$ correspond directly to the basis \eqref{eq:comp_basis} of the flip-flop coined QW)
	\STATE construct the continuous-time Hamiltonian on $G_{\text{exp}}$, as described in Sec.~\ref{sec:construction} (e.g. Eq.~\eqref{eq:hamil_example})
	\WHILE{$t<t_f$}
	\STATE \qquad $t \gets t+1$
	\STATE \qquad switch off all edges that intersect dashed curves (cf. Fig.~\ref{fig:graph_example})
	\STATE \qquad evolve the system for time $\frac\pi2$
	\STATE \qquad restore the edges that were switched off in line 7
	\STATE \qquad switch off all edges that do not intersect dashed curves
	\STATE \qquad evolve the system for time $\frac\pi2$
	\STATE \qquad restore the edges that were switched off in line 10
	\ENDWHILE
\end{algorithmic}
\end{algorithm}

\section{Conclusion}
\label{sec:conclusion}

Starting from a flip-flop coined QW on a generic graph $G$, we have described the construction of an expanded graph $G_{\text{exp}}$. This expanded graph serves two purposes. Firstly, it provides a visualization of the coin space, shedding new light on the coined QW model. Secondly, it allows the definition of a percolated CTQW on $G_{\text{exp}}$ that exactly reproduces the evolution of the original DTQW. Here, ``percolation'' means that edges are switched on and off in a systematic way during the continuous-time evolution, and it is needed to accommodate the alternating application of coin and shift operators in the coined model.

Our method shines when $G$ is regular and when Grover coins are used at all its vertices. In this case, we have the canonical choice for the Hamiltonian that drives the equivalent CTQW, namely the graph Laplacian of $G_{\text{exp}}$. During the evolution, this Hamiltonian changes implicitly through percolation of the underlying graph, i.e. it changes to the graph Laplacians of different subgraphs of $G_{\text{exp}}$. Without the assumptions on regularity and the coins that are used, certain components of the graph Laplacian have to be weighted and transformed to obtain the Hamiltonian for the CTQW. It would be preferable to avoid the use of percolation; however, our constructions are exact, and they hence show the limitations for reconciling the two models.

The existing literature on the relation of DTQWs and CTQWs consists mainly of studies that obtain CTQWs as a limit of DTQWs. Our work differs in that we produce the action of a coined QW on a graph $G$ by means of a CTQW on a larger graph. We hope that it provides insight and serves as a new starting point for studying the relation between DTQWs and CTQWs, and that our interpretation by means of expanded graphs inspires novel DTQW models. As a final remark, we point out that the connection between CTQWs and flip-flop coined QWs can be extended to other other DTQW models via work such as Ref.~\cite{Portugal2016}.

\section*{Acknowledgments}

P.P. would like to thank CNPq for its financial support (grant n.~400216/2014-0).
R.P. acknowledges financial support from Faperj (grant n.~E-26/102.350/2013) and CNPq (grants n.~303406/2015-1, 4741\-43/2013-9).


\end{document}